\documentclass[]{mn2e}
\usepackage[dvips]{graphicx,color}
\usepackage{amssymb}
\title{Periodic variability during the X-ray decline of 4U 1636-53}
\author[I. C. Shih]{I.C. Shih$^{1}$\thanks{E-mail:icshih@astro.soton.ac.uk}, A.J. Bird$^{1}$, P.A. Charles$^{1,2}$, R. Cornelisse$^{1}$, D. Tiramani$^{1}$  \\ 
$^1$School of Physics \& Astronomy, University of Southampton, Southampton, SO17 1BJ \\
$^2$South African Astronomical Observatory, P.O. Box 9, Observatory 7935, South
Africa}
\begin{document}

%\date{Accepted 2005. Received 2005; in original form 2005}

\pagerange{\pageref{firstpage}--\pageref{lastpage}} \pubyear{2005}

\maketitle

\label{firstpage}

\begin{abstract}
We report the onset of a large amplitude, statistically significant periodicity ($\sim$46 d) in the RXTE/ASM data of the prototype X-ray burster 4U 1636-53, the X-ray flux of which has been gradually declining over the last four years. This behaviour is remarkably similar to that observed in the neutron star LMXB KS 1731-260, which is a long-term transient.  We also report on an INTEGRAL/IBIS observation of 4U 1636-53 during its decline phase, and find that the hard X-ray flux (20-100 keV) indicates an apparent anti-correlation with soft X-rays (2-12 keV). We argue that 4U 1636-53 is transiting from activity to quiescence, as occurred in KS 1731-260. We also suggest that the variability during the X-ray decline is the result of an accretion rate variability related to the X-ray irradiation of the disc. 
\end{abstract}


\section{Introduction}
4U 1636-53 is a proto-typical low-mass X-ray binary (LMXB) which was one of the first to reveal type-I X-ray bursts (Swank et al. 1976), thereby demonstrating the presence of a neutron star compact object. Since its initial discovery (Willmore et al. 1974), 4U 1636-53 has been observed on numerous occasions as a persistent X-ray source, and although varying in flux by a factor of 2 to 3 (Hoffman, Lewin \& Doty 1977; Ohashi et al. 1982; Turner \& Breedon 1986; Hasinger \& van der Klis 1989), it has always been detectable by a wide variety of X-ray missions.

Extensive studies of 4U 1636-53 over a wide range of wavelengths (optical, UV and X-ray) have led to the establishment of some of its key parameters: (1) the orbital period of 3.8h; (2) the presence of a late-type, low-mass ($\sim0.4M_{\odot}$) donor transferring material onto the neutron star (Fujimoto \& Taam 1986, van Paradijs et al. 1990); (3) the presence of kHz QPOs (quasi-periodic oscillations) which are also seen during X-ray bursts, confirming that the neutron star has been spun-up (Zhang et al. 1996; Strohmayer 1999).  Some of these features are correlated with the mass accretion rate as they depend on X-ray intensity (Di Salvo, M\'{e}ndez \& van\ der\ Klis 2003).

All bright X-ray sources have been monitored daily since 1996 by the All Sky Monitor (ASM) onboard the Rossi X-ray Timing Explorer (RXTE).  We noted that the long-term lightcurve of 4U 1636-53 was exhibiting a gradual decline in X-ray flux since 2000, before which it had been persistent and stable at $\sim$20 ASM cts s$^{-1}$ in the 2-12 keV band.  More interestingly, it was clear that this decline was accompanied by a dramatic increase in the scale of its variability, and so we decided to investigate this behaviour in more detail, the results of which are presented here.  This behaviour appeared already to be very similar to that exhibited by another LMXB, KS 1731-260. Discovered in 1989 as an X-ray transient and type-I X-ray burster by MIR-KVANT (Sunyaev et al. 1990), KS 1731-260 subsequently remained ``on'' for more than 10 years.  However, Revnivtsev \& Sunyaev (2003, hereafter RS03) then reported the presence of a long-term ($\sim$38 d) quasi-periodic variability in the RXTE/ASM lightcurve of KS 1731-260, which appeared when the X-ray flux was declining during the interval 1998-2001 (and subsequently entering complete quiescence, see Wijnands et al. 2002).  RS03 concluded that this variability was unlikely to be the orbital period (which remains unknown for this object) and suggested that it was connected with the presence of a precessing accretion disc.

\section{Observations and Data Analysis}

\begin{figure*}
%\rotatebox{270}{\includegraphics[width=125mm]{4u1636asm.ps}}
%\input{1636_test}
\includegraphics[width=170mm]{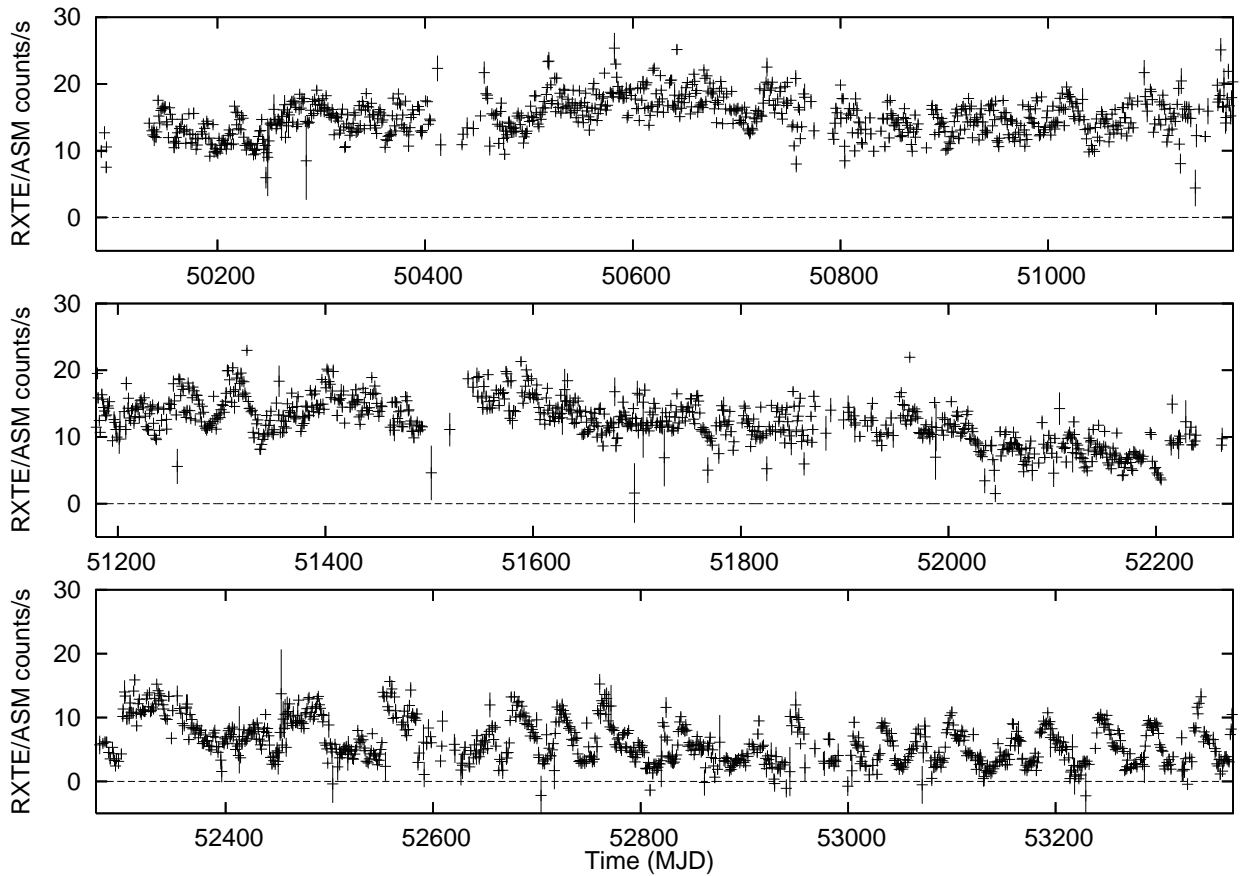}
%\vspace{3.5cm}
\caption{One-day average RXTE/ASM lightcurve of 4U 1636-53 from 1996 to  end of 2004. The dashed line represents zero flux and is shown for reason of clarity.}
\label{1636-536 asm lightcurve}
\end{figure*}

\subsection{RXTE/ASM}
The RXTE/ASM is designed to monitor the variable and unpredictable X-ray sky on a daily basis.  Since its launch in 1996, it has recorded the long-term intensity history of hundreds of bright X-ray sources. The ASM consists of three scanning shadow cameras (SSCs), each of which is sensitive in the energy range of approximately 1.5-12 keV, (and which can be sub-divided into the 1.5-3, 3-5 and 5-12 keV bands).  Each sky position is observed in a series of $\sim$90 s ``dwells'' several times per day (see Levine et al. 1996 and the MIT on-line archive http://xte.mit.edu for full details).  The ASM team at MIT maintains the data archive and updates it weekly.  Two forms of datasets are provided for all X-ray sources in their catalogue: one is the {\it individual dwells}, the other is the {\it one-day averages}. As we are interested in the variability on time-scales of days, we have used the one-day average data only.

The full ASM light curve of 4U 1636-53 from 1996 to the present day (MJD 50,088.1 - MJD 53,370, where MJD = JD - 2,400,000.5) is presented in Figure \ref{1636-536 asm lightcurve}. The X-ray time-history can be broken down into three main intervals:

\begin{enumerate}
\item 1996-1998 (MJD 50,088 - MJD 51,178): flux is stable and relatively bright ($\sim$15 cts s$^{-1}$) with low amplitude variability;
\item 1999-2001 (MJD 51,179 - MJD 52,274): flux is gradually declining to under 10 cts s$^{-1}$;
\item 2002-2004 (MJD 52,275 - MJD 53,370): flux is now low ($<$10 cts s$^{-1}$ ) and much stronger variability is clearly present.  At minimum, the source is only just detectable by the ASM (i.e. around 1 cts s$^{-1}$).
\end{enumerate}

\begin{figure}
%\rotatebox{270}{\resizebox{!}{85mm}{\includegraphics{4u1636asmlsp.ps}}}
%\rotatebox{270}{\resizebox{!}{85mm}{\includegraphics{4u1636asmlspthird.ps}}}
\includegraphics[width=85mm]{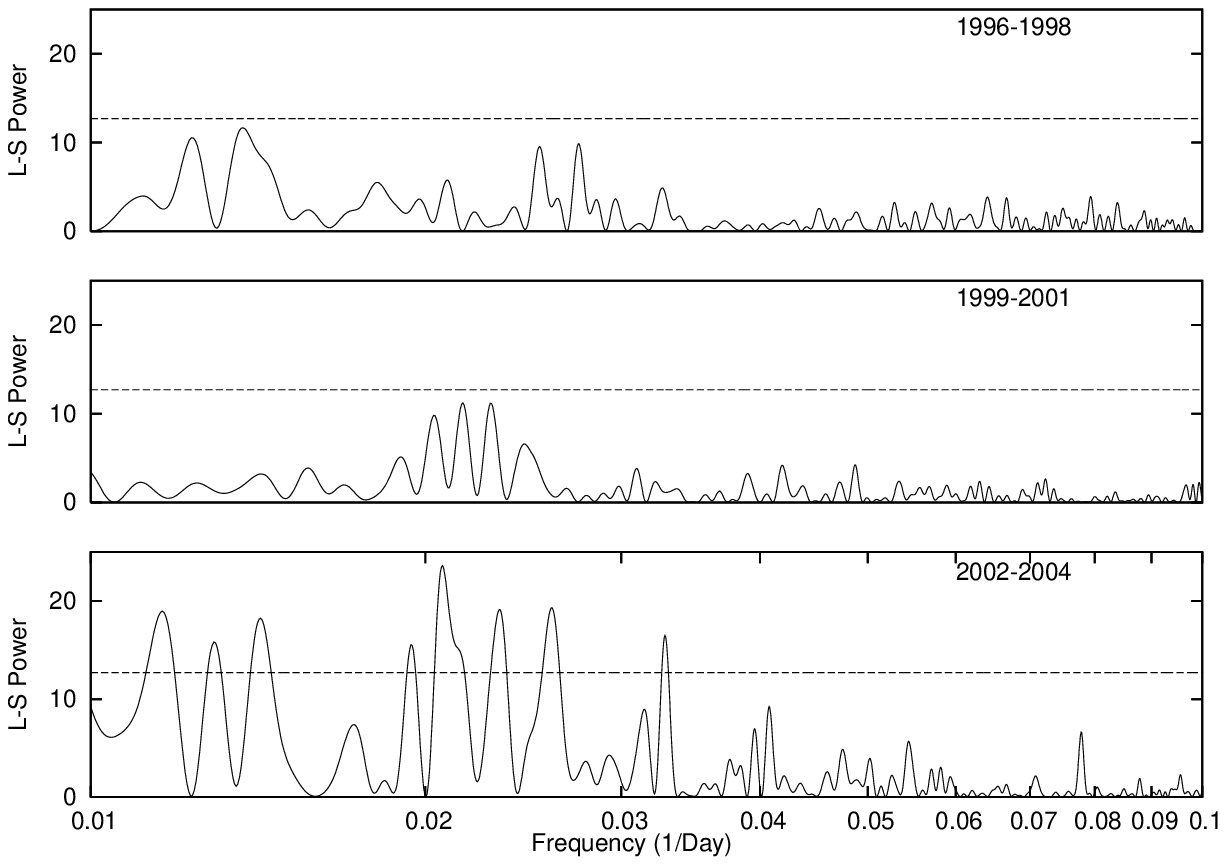}
\includegraphics[width=85mm]{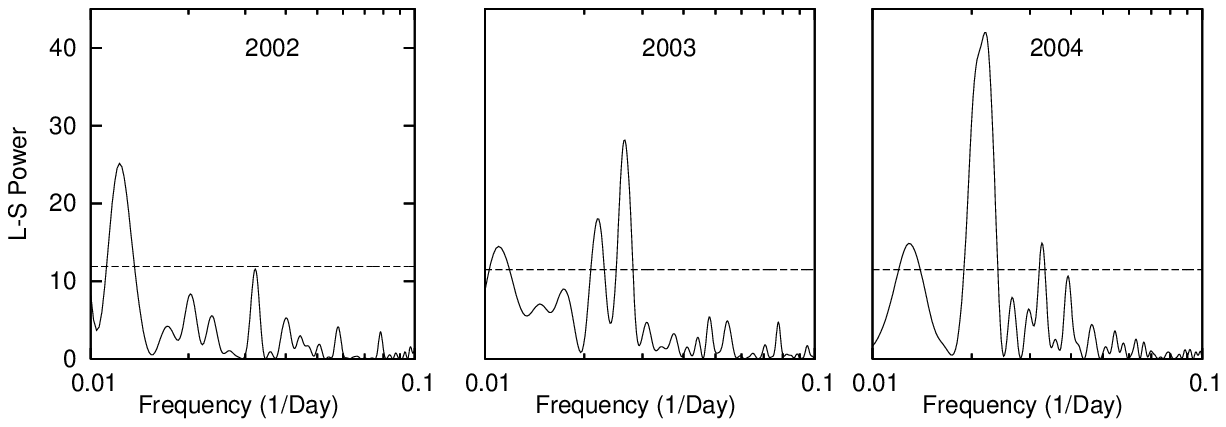}
%vspace{3.5cm}
\caption{Lomb-Scargle power spectra of 4U 1636-53 derived from the RXTE/ASM one-day average data using the default 1 d time bin, and the dashed lines represent the confidence level of $99\%$. The top three panels divide the 1996-2004 database into three approximately equal intervals. There are several significant signals shown in the 2002-4 interval and this is further subdivided in the lower panels.}
\label{lomb-scargle periodograms}
\end{figure}

To investigate the temporal properties of 4U 1636-53, we first linearly detrended the data, and then applied a fast fourier transform over the full ASM energy range of 2-12 keV.  We used the Lomb-Scargle method since it is very suitable for unevenly sampled time-series data such as that from the ASM (see Scargle 1982).  In Figure \ref{lomb-scargle periodograms} (upper panels) we present the resulting periodograms for each of the intervals identified above.
 
The X-ray flux in the first interval (1996-1998) shows no significant variability in the power spectrum.  In the second interval
(1999-2001), when the X-ray decline had begun around 2000, although subtle variations can be seen in both X-ray flux and the power spectrum, they are barely statistically significant.  However, in the most recent interval (2002-2004), well into the decline phase, not only is the variability very clear in the X-ray light-curve, but there are several highly significant periodic signals detected. The strongest signal represents a period of $\sim$46 days, with a confidence level of $99.99\%$ which is based on 10,000 times white noise simulations which assumed the same variance as the original RXTE/ASM data.

In fact, the presence of several significant signals in the 0.02-0.03 $d^{-1}$ frequency interval indicates that the variability in the X-ray decline of 4U 1636-53 is a {\it quasi}-periodic phenomenon. In addition, there is a very low frequency signal associated with the start of the decline. To examine the nature of this evolution in more detail we produced periodograms by subdivided the data onto an annual basis (see Figure \ref{lomb-scargle periodograms} (lower panels)). 

From these latter plots, we see the variability evolve to a slightly longer period with greater amplitude as the X-ray flux continues to decline. In 2003, the period was $39.4\pm0.7$ days, but in 2004 it was $46.7\pm4.4$ days. With these signals only present in the decline phase, this indicates that the variability is a function of the mass accretion rate onto the neutron star.
 
The variability we found is energy-independent in the ASM energy range of 2-12 keV (Figure \ref{energy band} (lower)), showing that the flux modulations of the three ASM sub-bands are identical.  This implies that the modulation is not due to the obscuring of a steady source (e.g. by varying amounts of absorbing material internal to the system), but is instead due to a modulation that is directly applied to the mass accretion rate. 

\subsection{INTEGRAL/IBIS}

Neutron star systems are known to produce hard, as well as soft, X-ray components (e.g. White et al. 1995), and so we took advantage of the large increase in $\gamma$-ray sensitivity afforded by INTEGRAL to obtain data on 4U 1636-53's behaviour at much higher energies.  We employed hard X-ray data on 4U 1636-53 from the IBIS/ISGRI instrument onboard the INTEGRAL $\gamma$-ray observatory (Ubertini et al. 1996). IBIS is a $\gamma$-ray imager operating in the energy range 20 keV to 10 MeV. The lightcurve was created using OSA version 4.1 provided by the Integral Science Data Centre (ISDC). All science windows when 4U 1636-53 was in the FOV of ISGRI were analysed, producing a light curve with typically 2000s time resolution, this being the typical pointing duration during the Galactic Plane Scans. Fluxes and errors were extracted directly at the known source position within each science window image, and the standard corrections for off-axis uniformity were applied.

\begin{figure}
%\rotatebox{270}{\resizebox{!}{85mm}{\includegraphics{4u1636energy_test.ps}}}
\includegraphics[width=85mm]{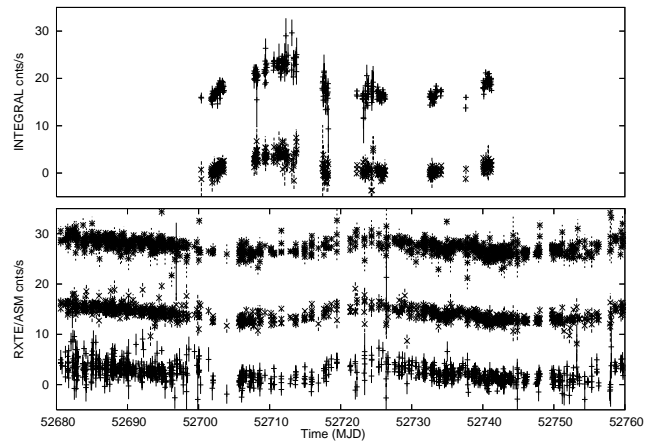}
%vspace{3.5cm}
\caption{Anti-correlation between RXTE/ASM and INTEGRAL fluxes. (Upper panel) INTEGRAL/IBIS light curves for two energy bands: 20-40 keV and 40-100 keV, the latter being shifted up by 15 cnt s$^{-1}$. (Lower panel) RXTE/ASM light curves for three energy bands: 1.5-3, 3-5, and 5-12 keV (bottom to top). For ease of comparison, the latter are also shifted up by 12 and 25 cnt s$^{-1}$ respectively. The RXTE/ASM data used here is the `` individual dwell" format.}
\label{energy band}
\end{figure}

For X-rays above  20 keV, we saw an unexpected anti-correlation with lower energy X-rays (see Figure \ref{energy band} (upper)). However, this is the only INTEGRAL dataset we are able to access at this time. Without additional data before and after the event, it is harder to interpret the significance of the apparent anti-correlation. It may be a time delay between soft  and hard X-rays, but this seems unlikely, given its duration, and the compact nature of the binary. It is more likely that the hard and soft components arise from different parts of the system.

\section{Discussion}

From a simple examination of the ASM lightcurve of 4U 1636-53, it is clear that the system is undergoing an X-ray state transition which began around 2000.  Such a transition is almost certainly directly related to a change in the average mass transfer rate from the donor star, and we believe this produces a subsequent change in the behaviour of the accretion disc that then modulates the effective accretion rate onto the neutron star. If this is the case, then why has such behaviour not been seen before? In fact, we believe it has, but it requires specific circumstances for such modulations to be detected and recognised.  In particular, the decline into quiescence has to be slow enough that the source is still detectable by the ASM for a long enough interval to establish a good baseline for periodicity searches.

This is where the comparison with the MIR/KVANT source KS 1731-260 is important, that was mentioned in section 1.  Already known to be a neutron star LMXB, RS03 also used the RXTE/ASM to discover a quasi-periodic modulation of  $37.67\pm0.03$ days. Similarly to 4U 1636-53, this was only seen during the decline phase which followed its almost 10 year ``on'' cycle (and which indicates one of the classification difficulties of X-ray binaries when the characteristic timescales of their behaviour are comparable to that of the history of X-ray astronomy!).  While we do not know the orbital period of KS 1731-260 (it is heavily obscured, so no optical identification exists), it is very likely to be short ($\sim$hours), and the optical modulation in decline is also very similar in amplitude to that of 4U 1636-53.  If 4U 1636-53 completes the mirroring of KS 1731-260 by declining into full quiescence, then it may indicate another characteristic of X-ray binary behaviour which is driven by a long-term drop in the mass transfer rate.

\begin{figure}
\includegraphics[width=85mm]{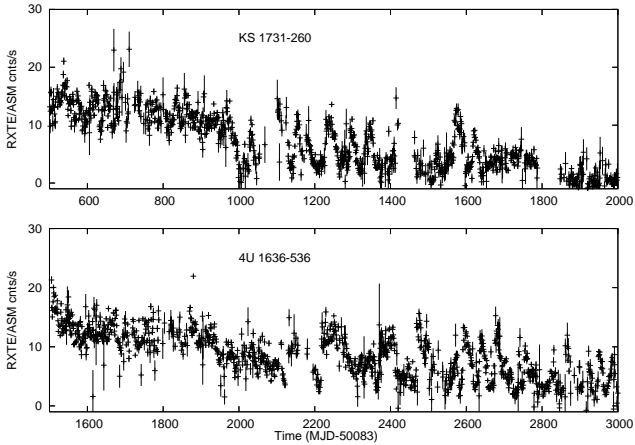}
%vspace{3.5cm}
\caption{Comparison of 4U 1636-53 and KS 1731-260 during their X-ray decline phases. KS 1731-260 has a shorter decline stage ($\sim800$ days) between the persistent and quiescent states, and the period discovered by RS03 is based on the interval from 1,100-1,500 days. In contrast, 4U 1636-53 has been in a more gradual X-ray decline for more than three years.}
\label{KS1731 and 4U1636}
\end{figure}

Obviously related systems include the ``classical'' X-ray transients themselves, sometimes referred to as {\it X-ray novae} (see e.g. McClintock \& Remillard 2003).  With recurrence timescales of decades, their outbursts normally last a few months as they decline exponentially into quiescence (although there is a wide range of lightcurve properties, see e.g. Chen, Shrader \& Livio 1997).  Whilst a key feature of the outburst lightcurve is the presence of a ``secondary maximum'', normally $\sim$60--90 days after the initial peak, many of them also display ``reflares'' later in the decline.  Examples of such lightcurves are collected in Kuulkers' (1998) paper (see his figure 8, 9), and demonstrate timescales for these reflares that are comparable to the quasi-periodicities seen in 4U 1636-53 and KS 1731-260 (see Figure \ref{KS1731 and 4U1636}), and consequently they may have a common physical origin.

We now consider what physical processes might produce modulations on these timescales.

\subsection{The effects of X-ray irradiation}
The viscosity in the accretion disc removes the angular momentum of the material transferred from the secondary star, thereby allowing it to be transported inwards to the central compact object. According to the classical ($\alpha$-disc) accretion disc model, the key factor in the behaviour of the accretion disc is the hydrogen ionisation temperature $T_{H}\sim6500$ K (Shakura \& Sunyaev 1973). The effective temperature at the local disc surface $T_{\mathrm{eff}}$ is related to the local mass density $\Sigma$, which is itself connected to the mass transfer rate $\dot{M}$. When the density increases sufficiently, $T_{\mathrm{eff}}$ can exceed $T_{H}$, and the disc becomes the hot, ionised with increased viscosity, thereby allowing accretion to take place. This is the well-known $T-\Sigma$ relation which is used to explain the long, low X-ray flux between brief outbursts in soft X-ray transients (SXTs) and those white dwarf analogues, the dwarf novae (Cannizzo 1993; King \& Ritter 1998).

However, for persistent LMXBs like 4U 1636-53, their X-ray luminosity is high ($\sim10^{37}$ erg s$^{-1}$; van Paradijs 1996), and the irradiation temperature $T_{irr}$ will dominate the local viscous temperature $T_{visc}$ at all accretion disc radii. Because L$_{X}\sim\eta\dot{M}c^2$, where $\eta$ is the efficiency of accretion, and as long as the mass accretion rate remains high enough, i.e. $\dot{M}\gtrsim10^{-10}M_{\odot}$ yr$^{-1}$, the X-ray irradiation alters the  $T-\Sigma$ relation by keeping the disc in a hot, high viscosity state and hence the source is continuously accreting (Dubus et al. 1999).   

The ``persistent'' description clearly indicates the observed nature of these bright LMXBs, and so we do not normally see large X-ray flux state changes as in the X-ray transients. So KS 1731-260 and 4U 1636-53 provide us with a rare opportunity to study the evolution of accretion discs in such circumstances (of slowly decreasing $\dot{M}$).

As the source gradually declines in X-rays (presumably a result of a decrease in the mass transferred from the secondary) the X-ray luminosity becomes too low to be able to ionise the entire accretion disc. This allows the outer region to cool down and thereby reduce the overall mass accretion rate onto the compact object, subsequently leading to an X-ray minimum. Meanwhile, the disc continues to accrete from the secondary star and, through the normal $T-\Sigma$ relation, will eventually re-enter the hot, viscous state, thereby re-starting X-ray activity.

\begin{figure}
%\rotatebox{270}{\resizebox{!}{85mm}{\includegraphics{4u1636timescale.ps}}}
\includegraphics[width=85mm]{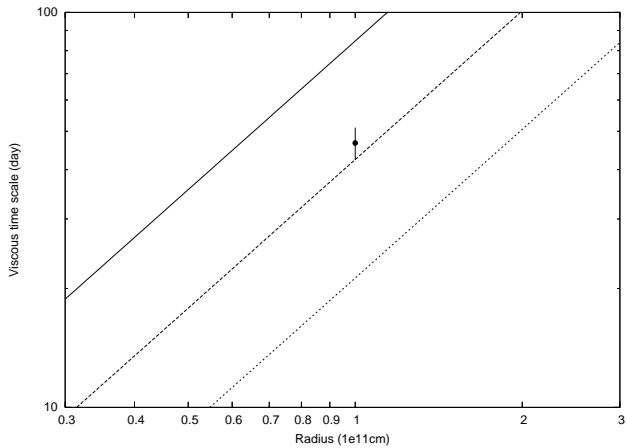}
%vspace{3.5cm}
\caption{Viscous time scales for a Shakura-Sunyaev $\alpha$-disc as estimated for the parameters of 4U 1636-53 and assuming $\alpha$-disc with $\alpha=0.1$ and $M_{1}=1.4M_{\odot}$. The solid point marks the observed period of 46.7$\pm$4.4 day at the suspected outer disc radius. The solid, dashed, and dotted lines correspond to $\dot{M}$ values of $10^{-12}, 10^{-11}$, and $10^{-10} M_{\odot}$ yr$^{-1}$ respectively.}
\label{time scale}
\end{figure}

This would imply a timescale for this process close to the viscous time scale in the outer disc. Following Truss et al. (2002)  we estimate this timescale by assuming that the outer disc is a Shakura-Sunyaev $\alpha$-disc, with viscous timescale given by

\begin{equation}
t_{v}\sim\frac{R^2}{\alpha c_{s}H}.
\end{equation} which Truss et al. parameterise using a Kramer opacity law as

\begin{equation}
t_{v}\sim 390\left(\frac{\alpha}{0.1}\right)^{-4/5}\dot{M}_{16}^{-3/10}M_{1}^{1/4}R_{11}^{5/4} d
\end{equation} where $\dot{M}_{16}$ is the mass accretion rate in units of $10^{16}$ g s$^{-1}$, $M_{1}$ is the accreting object mass in solar masses and $R_{11}$ is the disc radius in units of $10^{11}$ cm. Figure \ref{time scale} plots $t_{v}$ for the parameters of 4U 1636-53, and indicates that $t_{v}$ at the outer edge of the disc ($\sim10^{11}$ cm) is roughly comparable to the rebrightening timescale of 4U 1636-53 ($40\sim50$ days).

Furthermore, while the outer part of the disc is cool and not accreting, the inner region will likely be replaced by a hot thin corona because the high density disc material here has been exhausted. This would then explain the observed anti-correlation between the soft and hard X-ray fluxes in 4U 1636-53. We also note that the anti-correlation between the hard and soft X-rays happens at a L$_{X}\sim 5\%$ of the Eddington limit, a comparable level where black hole transients show a spectral state transition between the low/hard and the high/soft states (McClintock \& Remillard 2003). For example, Cadolle Bel et al. (2004) observed this transition in the black hole candidate XTE J1720-318 with XMM-Newton, RXTE and INTEGRAL which provide a truly broad range of energy spectrum (1-200 keV). This suggest that the behaviour of 4U 1636-536 at low accretion rates is comparable to that of the black hole candidates.

\subsection{Tidal effect}
Short period (few hours or less) interacting binaries must have a low mass ratio $q\lesssim1/3$ (where $q=M_{2}/M_{X}$)  and this means that the outer edge of the accretion disc is close to the tidal radius, or Roche lobe of the primary. Numerical simulations (e.g. Whitehurst 1988; Whitehurst \& King 1991) have shown that, in such low $q$ systems, it is possible for the disc to reach this radius , thereby allowing tidal forces (due to the secondary) to trigger the 3:1 resonance in the disc.  

Given the parameters of 4U 1636-53, it has a mass ratio of $q\sim$0.33. and so it is possible for the outer edge of the accretion disc to be influenced by the tidal forces of the secondary star. It would then develop into an eccentric disc and undergo enhanced viscous dissipation, as demonstrated by Truss et al. (2002) who used an SPH simulation to account for the rebrightening in the lightcurve of A0620-00.

Whilst this process should be occuring in 4U 1636-53 (and was used by RS03 to account for the observed modulation in KS 1731-260), our INTEGRAL observations indicate that a more complex explanation (such as X-ray irradiation as given in section 3.1) is required.

\section{Conclusion}
It is clear that 4U 1636-53 is undergoing an X-ray state transition from the bright, active phase to that of quiescence as a result of a decrease in the overall mass accretion rate. This is presumably linked to a decrease in the mass transfer rate into the disc, and this has triggered a change in its properties that then modulates the accretion rate onto the neutron star. We suggest that standard  $\alpha-disc$ disc instability model can account for the variability timescales observed in the X-ray decline of 4U 1636-53, providing the effects of X-ray irradiation are included. The long-term quasi-periodicities observed in the decline phases of 4U 1636-53 and KS 1731-260 may be more widely present in LMXBs and X-ray transients than had previously been recognised. This suggests that it is a key property of LMXB accretion discs in their mid to low accretion rate states.
 
\section{Acknowledgements}
We would like to thank the ASM/RXTE teams at MIT and GSFC for provision of the ASM data. The data used in this paper include the observations of INTEGRAL, an ESA project with instruments and science data centre funded by ESA member states (especially the PI countries: Demark, France, Germany, Italy, Switzerland, and Spain), the Czech Republic and Poland with the participation of Russia and the USA.

\end{document}